# The effect of dataset size and the process of big data mining for investigating solar-thermal desalination by using machine learning

Guilong Peng[#1], Senshan Sun[#2], Zhenwei Xu[2], Juxin Du[2], Yangjun Qin[2], Swellam W. Sharshir[3], A.W. Kandel[3], A.E. Kabeel[4,5], Nuo Yang*[6]

[1]School of Mechanical and Energy Engineering, Shaoyang University, Shaoyang 422000, China

[2]School of Energy and Power Engineering, Huazhong University of Science and Technology, Wuhan 430074, China

[3]Mechanical Engineering Department, Faculty of Engineering, Kafrelsheikh University, Kafrelsheikh 33516, Egypt

[4]Mechanical Power Engineering Department, Faculty of Engineering, Tanta University, Tanta, Egypt

[5]Faculty of Engineering, Delta University for Science and Technology, Gamasa, Egypt

[6]Department of Physics, National University of Defense Technology, Changsha 410073, China

#Guilong Peng and Senshan Sun contribute equally to this work

*Corresponding email: Nuo Yang (nuo@hust.edu.cn, nuo@nudt.edu.cn)

## Highlights

- The proposed platform saves more than 83% of the time for data collection.
- Over 1000 experimental datasets were collected for machine learning analysis.
- Three different algorithms were compared from several important aspects.
- A standard process flow of the interdisciplinary study is proposed.

# Abstract

Machine learning's application in solar-thermal desalination is limited by data shortage and inconsistent analysis. This study develops an optimized dataset collection and analysis process for the representative solar still. By ultra-hydrophilic treatment on the condensation cover, the dataset collection process reduces the collection time by 83.3%. Over 1,000 datasets are collected, which is nearly one order of magnitude larger than up-to-date works. Then, a new interdisciplinary process flow is proposed. Some meaningful results are obtained that were not addressed by previous studies. It is found that Radom Forest might be a better choice for datasets larger than 1,000 due to both high accuracy and fast speed. Besides, the dataset range affects the quantified importance (weighted value) of factors significantly, with up to a 115% increment. Moreover, the results show that machine learning has a high accuracy on the extrapolation prediction of productivity, where the minimum mean relative prediction error is just around 4%. The results of this work not only show the necessity of the dataset characteristics' effect but also provide a standard process for studying solar-thermal desalination by machine learning, which would pave the way for interdisciplinary study.

# Nomenclature

| | |
|---|---|
| AF | Formula of activation function |
| b | Bias in BP-ANN |
| $\widehat{B}$ | List of regression coefficients |
| $c_1$ | Average values of productivity in $R_1(j)$ in RF |
| $c_2$ | Average values of productivity in $R_2(j)$ in RF |
| D | Normalized dataset |
| $e$ | Label of node |
| $E$ | Output error signal in BP-ANN |
| $E_{min}$ | Threshold of root mean square error in BP-ANN |
| $f_i$ | Predicted value in ML |
| $H_F$ | Fan height above the basin |
| $GI$ | Gini impurity |
| $h$ | Hyperparameters of BP-ANN |
| i | Label of neurons in current layer in BP-ANN |
| j | Label of sample in RF |
| k | Number of independent variables |
| l | Label of neurons in previous layer in BP-ANN |
| m | Number of neurons in current layer in BP-ANN |
| $M$ | label of regions in RF |
| $\dot{m}$ | Productivity |
| $m_t$ | Total mass of the collected freshwater increases with time |
| n | Number of DTs in RF |
| N | Sample size |
| $N_M$ | Number of elements in region $M$ in RF |
| $o_M$ | Average output value in DT |
| $\hat{p}_e$ | Estimated probability that sample belongs to any class at node $e$ in RF |
| $P_F$ | Power of the fan |
| $q$ | Dimension label |
| $R(j)$ | Regions sliced by $j_{th}$ sample |
| $R_M(j)$ | Region of label $M$ in RF |
| $R_l$ | Random number |
| $R^2$ | Coefficient of determination |
| $\Delta t$ | Given period |
| $T_{amb}$ | Ambient temperature |
| $T_g$ | Glass cover temperature |
| $T_{ss}$ | Solar still types |
| $T_w$ | Water temperature |
| $VIM_i^{DT}$ | Importance of one variable at node $e$ in RF |
| $w$ | Weight between current layer and previous layer |
| $x$ | Input value |
| X | Array of experimental independent variables |
| $y$ | True productivity |
| $\bar{y}$ | Average productivity of datasets |
| $y^{MLR}$ | Predicted value in MLR |
| $y^{neu}$ | Output of current neurons |
| $y^{NN}$ | Predicted value in BP-ANN |
| $y^{RF}$ | Predicted value in RF |
| $y_i$ | One of productivity |
| Y | List of productivity |

## Greek letters

| | |
|---|---|
| $\alpha_q$ | Value of $q_{th}$ neuron in first layer |
| $\beta$ | Regression coefficient |
| $\beta_h$ | Value of $h_{th}$ neuron in second layer |
| $\delta$ | relative prediction error |
| $\bar{\delta}$ | mean relative prediction error |
| $\delta^e$ | Error signal between current layer and previous layer |
| $\eta$ | Scale coefficient |

## Abbreviations

| | |
|---|---|
| ANN | Artificial neural network |
| BO | Bayesian optimization |
| BP-ANN | Backpropagation artificial neural network |
| CART | Classification and regression trees |
| DT | Decision tree |
| ML | Machine learning |
| MLR | Multiple linear regressions |
| RF | Random forest |
| STD | Solar-thermal desalination |
| XPS | Extruded polystyrene |

## 1. Introduction

The problem of safe drinking water is becoming increasingly serious due to the unbalanced distribution of water resources and environmental pollution, which leads to many problems, especially the health problems of residents in underdeveloped areas [1, 2]. Seawater desalination technology plays an important role in solving this problem [3]. Among the seawater desalination technologies, solar-thermal desalination (STD) has drawn much attention in the last decades[4, 5], especially for small-scale and micro-scale systems[6, 7], because of its simplicity, low investment cost, portability, and so on [8, 9].

Accurate productivity prediction and factor analysis are important to STD, which not only help to evaluate their practical potential but also provide guidance for future optimization [10]. Conventional productivity prediction and factor analysis rely on a physics-based modeling approach, which provides a fundamental understanding of the physics process. Nevertheless, it is difficult to accurately predict or analyze a practical system with only a physics-based model, because many factors cannot be easily depicted by a physics-based model, hence many assumptions or half-empirical correlations are needed, such as the factors that cannot be quantitatively described or is correlated indirectly[11]. Thus, physics-based models are good only for very simple systems.

On the other hand, machine learning (ML) methods provide a prediction and analysis approach regardless of the complexity of the system. Therefore, ML has attracted much attention in many scientific fields, such as chemistry, physics, materials science, biology, and so on [12, 13], because of its advantages in massive data analysis capabilities, saving labor, economic costs, and time. The application of ML methods has been well-studied in many solar energy fields, such as solar thermal collectors [14], solar cells [15], and solar radiation forecasting [16]. Therefore, an interdisciplinary study between ML and STD may also have great potential.

Various algorithms have been applied to fit the results of STD systems in the past decades, including artificial neural networks (ANNs) [17], random forest (RF)[18, 19], hybrid fuzzy-neural algorithms [20], modified krill herd (MKH) algorithm [21], modified random vector functional link (RVFL) [22], and so on. For example, Noe et al. [23] found that up to 89% of the predictions were within 20% of actual production by using ANN based on 312 datasets. Mashaly et al. [24] developed a back propagation ANN model for the prediction of solar still performance, with a coefficient of determination ($R^2$) of 0.93 for predicting the productivity of seawater desalination. Later, they compared multi-layer perceptron neural networks and multiple linear regressions. The results showed that the average value of $R^2$ for the multi-layer perceptron model was higher by 11.23% than for the multiple linear regressions model [25]. Recently, more efforts have been made to further optimize the conventional algorithms, such as developing the Imperialist Competition Algorithm enhanced ANN algorithm [26], optimizing ANN by using Harris Hawks Optimizer [27, 28] and Levenberg Marquardt algorithm [29, 30], and so on. The best-reported $R^2$ reaches up to nearly 1 [31, 32].

However, most previous investigations only focused on enhancing the fitting accuracy of a given dataset, such as increasing $R^2$ or decreasing the relative errors of productivity prediction[33]. The application of ML is quite limited and far from becoming an essential tool. One of the reasons is the insufficient dataset size, usually less than 200 due to the time-consuming data collection process [23, 34]. Such a small dataset makes it impossible to carry out more discussion. On the other hand, the ML analysis conditions vary across different works, leading to difficulties in drawing systematic conclusions. For instance, the dataset sizes and ranges differ in various works, making direct comparisons challenging [35]. Therefore, it is vital to propose a general and reasonable process to enrich the results and make them more comparable across different works.

The main objectives of this work are: (1) to explore a representative new method for speeding up and expanding the dataset collection of STD systems; (2) to propose an optimized standard process flow for analyzing STD systems by the ML method, which tries to eliminate the inconsistency across different works as much as possible; (3) to explore more possibility of interdisciplinary study that beyond the limitation of conventional fitting.

In this work, firstly, the representative STD systems, a few solar stills, were designed and optimized for collecting a large dataset of productivity, temperatures, and other factors. Then, it was proposed that a standard process flow of analyzing STD systems by using ML, which consists of seven steps. Lastly, several important aspects of the interdisciplinary study were explored, such as the effect of the dataset size on the prediction accuracy of productivity, the effect of the dataset range on the importance analysis of various influence factors, and the performance of productivity prediction by extrapolation.

## 2. Experimental platform and ML algorithms

Solar still, which is a typical small-scale STD system, is built as an example of investigating dataset collection and ML analysis. The solar still is a simple, low-cost, micro-scale STD system that requires minimal maintenance and has been extensively researched in recent decades[36, 37]. The dataset collection involves three different types of solar stills: single-slope, double-slope, and pyramid, which are the most popular types of solar still [38]. The basic working principle of a solar still is as follows: the seawater in the basin is heated by solar radiation and evaporates. The vapor rises due to natural or forced convection and condenses on the glass cover, which is cooler than the vapor [39]. The condensate then slides down the glass cover by gravity and is collected in bottles. For more details on solar stills, please refer to [40, 41].

Based on the experimental platform (Fig. 1a and 1b) and the thermodynamic processes of solar stills [42, 43], potential features affecting productivity can be identified and utilized as inputs for machine learning models. Three main types of factors can be considered: (1) temperature factors, including water temperature ($T_w$), glass cover temperature ($T_g$), and ambient temperature ($T_{amb}$); (2) air convection factors, including fan power ($P_F$) and fan height above the basin ($H_F$); and (3) geometry factor, i.e., the type of solar still ($T_{ss}$). $T_w$ and $T_g$ primarily depend on the heating power of the system. $T_{amb}$ is regulated by a thermostat cover above the glass cover, which can vary between 10°C and 35°C. Validation of the thermostat cover can be found in previous work [43]. A fan is mounted in the vapor chamber of the solar still, and $P_F$ and $H_F$ can be adjusted to investigate the impact of air convection within the solar still. The bottom of the solar still measures 25 cm × 25 cm and is insulated with a 4 cm layer of XPS (extruded polystyrene) foam. Additionally, a random number list, $R_l$, is generated by the computer for use as a reference. The list of devices used is presented in Table 1.

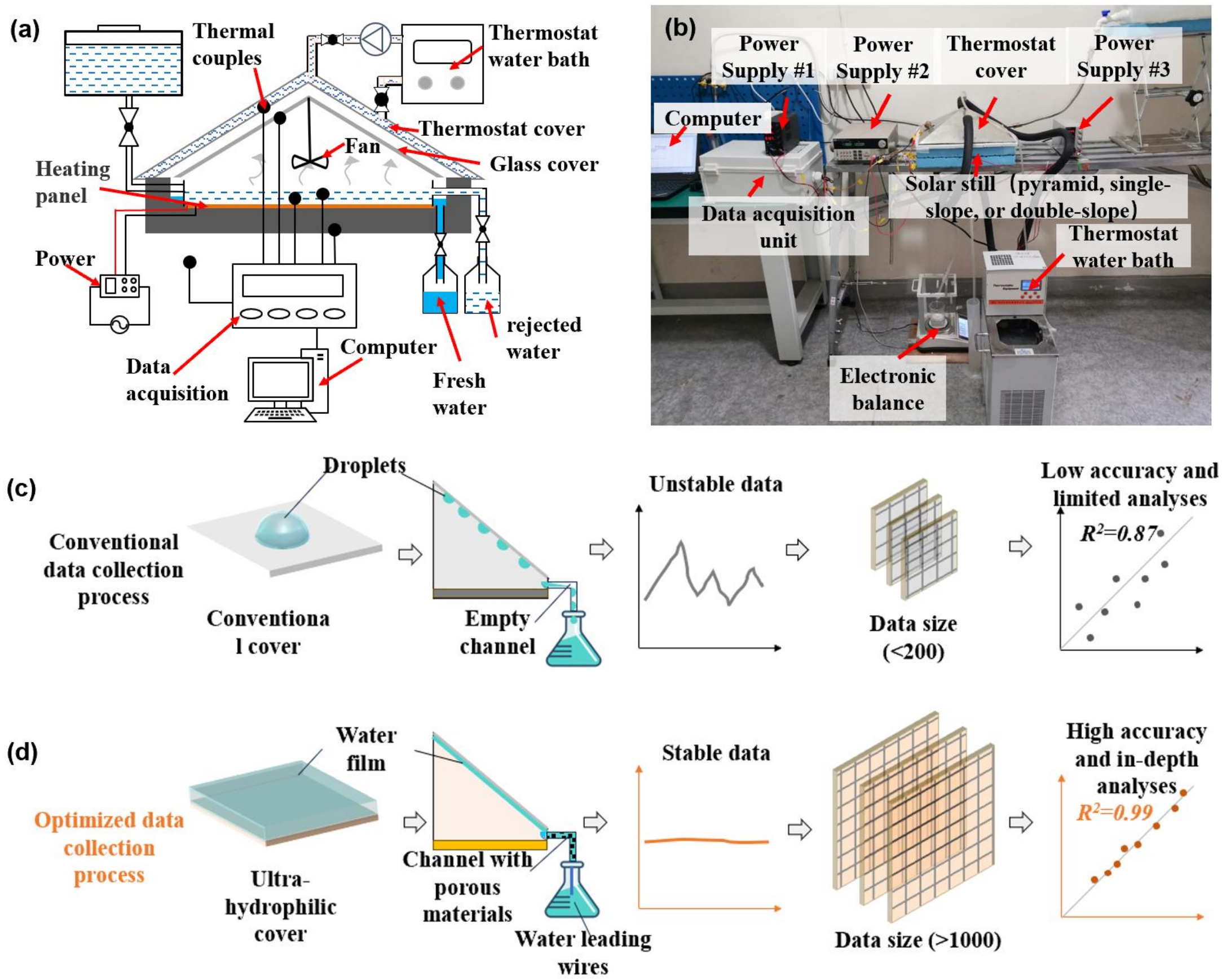

Fig.1 (a) Schematic diagram of the experimental platform (b) Photo of the experimental platform. Pyramid, double slope, and single slope solar still are tested in turn in the setup by substituting the cover of solar still (c) Schematic diagram of the conventional and (d) optimized data collection processes.

It is necessary to control each factor precisely to optimize the experimental procedures and investigate the effect of each factor. Therefore, the stable artificial environment, instead of the real environment, is used in this work. Herein, electrical heating is used for simulating solar energy, which provides stable power input and shows the steady-state performance of the system [44]. The power density of electrical heating ranges from 0 W/m$^2$ to 1,000 W/m$^2$. Besides, the ambient temperature is stably controlled by the thermostat cover as aforementioned. Therefore, the data range can be easily controlled as compared to conventional platforms.

Table 1 Specifics of devices and sensors in the experiments.

| Name | Brand | Type | Function | Range | Error |
|---|---|---|---|---|---|
| Fan | LFFAN | LFS0512SL | Enhancing convection | 0 ~ 4800 RPM | - |
| Electronic balance | ANHENG | AH-A503 | Measuring productivity | 0 ~ 500 g | ±0.01 g |
| Power supply #1 & #3 | WANPTEK | NPS3010W | DC power supply | 0 ~ 30 V | ±0.1 % |
| Power supply #2 | ITECH | IT6932A | Programmable power supply | 0 ~ 60 V | ± 0.03 % |
| Data acquisition unit | CAMPBELL SCIENTIFIC | CR1000X& AM25T | Dataset collection | 25 Channels | - |
| Thermostat water bath | QIWEI | DHC-2005-A | Controlling the ambient temperature | -20 ~ 99.9 °C | ± 0.2 °C |
| Heating panel | BEISITE | Custom-made | Heating the water | 0 ~ 2000 W/m$^2$ | - |
| Thermal couple | ETA | T-K-36-SLE | Measuring the temperature | -200 ~ 260 °C | ± 1.1 °C |

In a conventional solar still system, the freshwater condenses as droplets as shown in Fig.1c. The total mass of the collected freshwater increases with time ($m_t$). The instantaneous freshwater productivity ($\dot{m}$) can be obtained from the freshwater mass difference ($m_{t+\Delta t}-m_t=\Delta m$) during a given period ($\Delta t$), i.e., $\dot{m}=\Delta m/\Delta t$ . However, $\Delta m$ fluctuates with time, thus $\dot{m}$ would be intrinsically unstable, especially under a small period as shown in Fig. 2a. When the period is 5 mins ($\Delta t=5\ mins$), the instantaneous productivity in conventional solar still ranges from 23 g/h to 43 g/h, although the thermal equilibrium state has been reached. The productivity fluctuates by 70% around the average value, thus completely unstable. For $\Delta t$ = 15 mins, the productivity ranges from 27 g/h to 37 g/h, and the fluctuation remains as high as 18.5%. The fluctuation decreases to 10% when $\Delta t$ = 30 mins. Therefore, in a conventional system, a long collecting time is inevitable to obtain just one single reliable dataset.

After a careful evaluation of the data collection process, it was found that the unstable productivity of the conventional system resulted from the fluctuating falling frequency and size of freshwater droplets. To solve this problem, the glass cover is treated to be ultra-hydrophilic, which enables film condensation and avoids the unstable droplets in previous works. In this work, anti-fog coating (Rain-X, Illinois Tools Works Inc.) is used for ultra-hydrophilic treatment. More details about the ultra-hydrophilic glass cover can be found in our previous work [40]. The glass cover is recoated every week, due to the surface may exhibit degradation. The stability is verified every day before and after the experiments. Besides, the condensate from the ultra-hydrophilic glass cover flows continuously to a bottle through a fibrous water channel and a water-leading wire, as shown in Fig 1c. The weight of the collected condensate is recorded by the electronic balance every 10 seconds. After optimization, the instantaneous productivity ranges from 29.5 g/h to 33.5 g/h when $\Delta t$ = 5 mins, which fluctuates by only 7%. It is even better than that of $\Delta t$ = 30 mins in conventional solar still of previous works. Thus, the dataset collection time is saved by around 83.3% as compared to conventional systems, from 30 mins to 5 mins. Fig. 2b compares the standardized normal productivity distribution of different conditions. The productivity of the proposed system (film-wise) at $\Delta t$ = 5 mins is much more stable than that of the conventional system (drop-wise) even when its $\Delta t$ is as long as 30 mins.

Due to the significantly reduced dataset collection time, more datasets can be collected. Massive datasets are collected by changing the experimental condition. For example, datasets can be obtained automatically and continuously by changing the fan power (Fig. 2c). The stepwise fan power is controlled by the programable power supply. The corresponding productivity and temperatures in the stable state are recorded in the computer for further analysis. Then, one set of data that includes $\dot{m}$, $T_w$, $T_g$, $T_{amb}$, $P_F$, $H_F$, and $T_{ss}$ are successfully obtained. In this work, 1022 data were collected for analysis, which is ten times greater than the average number of datasets in previous works as shown in Fig. 2d. The diversity, integrality, and representativeness were considered when collecting the data. For example, the typical working temperature range for solar stills is from 10°C to 80°C, therefore the datasets cover all the typical temperatures. The difference between the conventional experimental platforms and the optimized platform in this work is shown in Table 2.

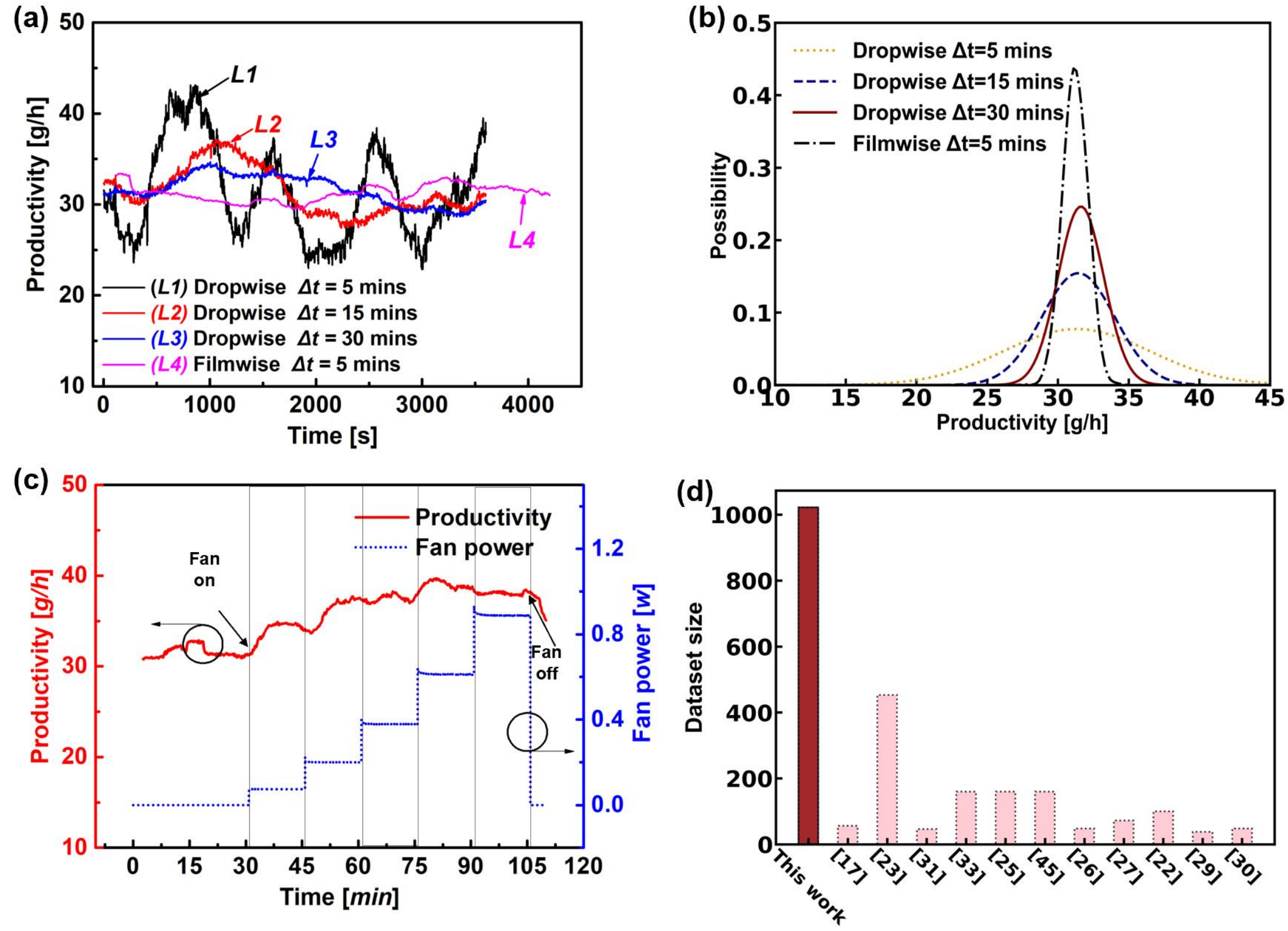

Fig. 2 (a) Hourly productivity under different periods. (b) Standardized normal distribution of productivity. (c) The productivity of solar still with stepped fan power. (d) The dataset size collected from the solar still in this work and the references.

Table 2 Comparison of the conventional platform and optimized platform

| Specifications | Conventional [22, 27, 45] | Optimized |
| --- | --- | --- |
| Collection time for a dataset | > 30 mins | 5 mins |
| Ambient temperature range | Uncontrolled | Controlled |
| Heating power range | Uncontrolled | Controlled |
| Fan power control | Manual | Automatic |
| Glass cover wettability | Hydrophilic | Ultra-hydrophilic |
| Water collection pipe | Empty pipe | A pipe full of porous materials |
| Freshwater flow | Droplets | Stream |

| Dataset size | <200 | 1,022 |
|---|---|---|

To analyze the dataset from the solar still, three different algorithms of ML were used and compared, including multiple linear regressions (MLR), backpropagation artificial neural network (BP-ANN), and random forest (RF).

**(1) MLR model**

MLR is an algorithm based on the least squares method that has been widely used in STD[25, 46]. It has the advantages of speediness and convenience. MLR is derived through the utilization of the least squares method, which aims to minimize the sum of squared residuals. The equation for multiple linear regression can be expressed in matrix form as follows

$$y^{MLR} = \hat{\beta}_0 + \hat{\beta}_1 x_1 + \hat{\beta}_2 x_2 + \cdots + \hat{\beta}_k x_k \quad (1)$$

$$X^T X \hat{B} = X^T Y \quad (2)$$

where $y^{MLR}$ is the predicted value, $x$ is the input value, $\beta$ is the regression coefficient. In the MLR model, the input dataset X is an array containing several experimental independent variables, such as $T_w$, $T_g$, $T_{amb}$, $P_F$, $H_F$, and $T_{ss}$. Y represents the list of productivity. The array $\widehat{B}$ represents the list of regression coefficients, which are determined through the fitting process using the dataset. Once $\widehat{B}$ has been fitted using the dataset, the productivity can be predicted using the MLR model. The detailed calculation processes can be found in “Supporting Information S3”. For the approaches, theoretically, they can be used for all physical models. However, some considerations or limitations should be noted, details in Supporting Information S4 to S6.

**(2) BP-ANN model**

ANN is a popular neural algorithm in the field of STD due to its high accuracy [47]. ANN belongs to the black-box model, and the specific formulas cannot be obtained. The principle of ANN is similar to the information transmission of biological neurons. BP-ANN is a type of ANN in which the signal is forward propagated and the error is backpropagated. The model is continually revised through continuous error feedback. Then a high-precision predicting model is obtained. The variables in BP-ANN are the values of neurons in the input layer. In this work, the number of neurons in the input layer is 6, corresponding to 6 independent variables of experiments, which serve as the input features of the neural network. By calculating the values of neurons in the hidden layer with their corresponding weights, the final result is obtained as the value of the output layer that has only one node. This value represents the predicted productivity based on the given independent variables and the BP-ANN model.

A five-layer perceptron with 3 hidden layers was used in this work. When a training sample is input, the output error signal E, of the BP-ANN model is

$$E=\frac{1}{2}\left(y^{NN}-y\right)^{2} \tag{3}$$

where $y^{NN}$ and $y$ are the predicting and true values.

Utilizing the BP-ANN algorithm, the model propagates the error signal back through the network and adjusts the weights between nodes, and the resulting updated outcome is represented as

$$w'=w+\Delta w=w-\frac{\partial E}{\partial w} \tag{4}$$

where $w$ is the weight between the current layer and the previous layer according to the backpropagation. The activation function employed in this study is the unipolar sigmoid function. The weight adjustment is performed according to

$$\Delta w_{il}=\eta\delta_{l}^{e}y_{i}^{neu} \tag{5}$$

where i is the label of the neurons in the current layer, and $l$ is the label of the neurons in the previous layer. $y^{neu}$ is the output of the neuron using the activation function in the current layer; $\eta \in (0,1)$ is the scale coefficient, a larger η means a faster convergence speed but the local optimum may not be obtained, and a smaller η means higher accuracy and slower convergence speed. $\delta$ is the error signal between the current neural network layer and the previous layer, the results can be calculated by reverse iteration (Eq. S20).

The bias can be represented as

$$b' = b + \Delta b = b - \frac{\partial E}{\partial b} \tag{6}$$

where b is the bias between the current layer and the previous layer according to the backpropagation. The bias is performed according to

$$\Delta b_j = \eta \sum_{j=1}^{m} \delta_j \tag{7}$$

For the training set with a sample size of N, the root mean square error is used as the total error of the model,

$$E_{RME} = \sqrt{\frac{1}{N} \sum_{i=1}^{N} \left[ \frac{1}{2} (y_i^{NN} - y_i)^2 \right]} \tag{8}$$

When $E_{RME} < E_{min}$ is satisfied or the maximum number of iterations is reached, the training ends, and the artificial neural network prediction model is obtained. The forward pass prediction process between the first and second layers is

$$\beta_h = AF \left[ \sum_{i=1}^{m_1} (w_{ih}^2 x_i - b_h^1) \right] \quad h = 1, 2, \cdots, m_2 \tag{9}$$

where $m_1$ is the number of neurons in the first layer, $\beta_h$ is the value of the $h_{th}$ neuron in the second layer, $\alpha_q$ is the value of the $q_{th}$ neuron in the first layer. AF is the activation function. A similar process happens between the second and third layers, as shown in Fig. 3. All neurons are computed in the forward process, resulting in the final output. Please refer to the "Supplementary Information S3" for detailed calculation procedures.

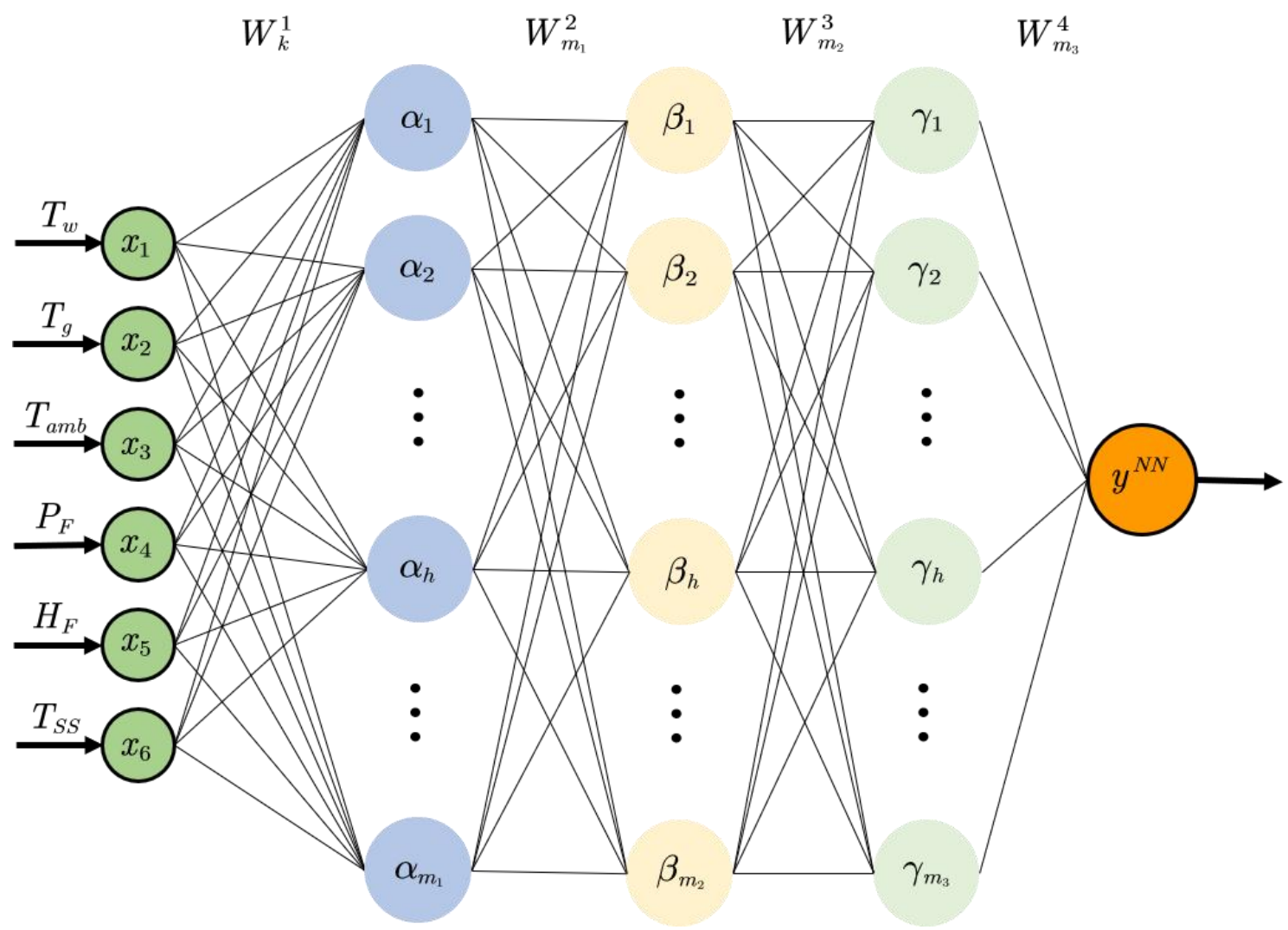

Fig. 3 Schematic diagram of BP-ANN algorithm.

The overall flow chart of using BP-ANN for predicting the productivity of the STD system is shown in Fig. 4. The initial input consists of over 1,000 experimental datasets that comprise six independent variables and one dependent variable, $\dot{m}$. To ensure consistent sample spacing, the initial dataset undergoes a data normalization process (Supporting Information S1). The initial BP-ANN model solely comprises BP-ANN algorithms, which are unable to make predictions before training. Bayesian optimization (BO) is employed to modify the BP-ANN model, providing optimized hyperparameters such as 'hidden_layer_sizes', 'Activation', and 'Solver' [48, 49]. For more details, please refer to the “Supporting Information Table S1”.

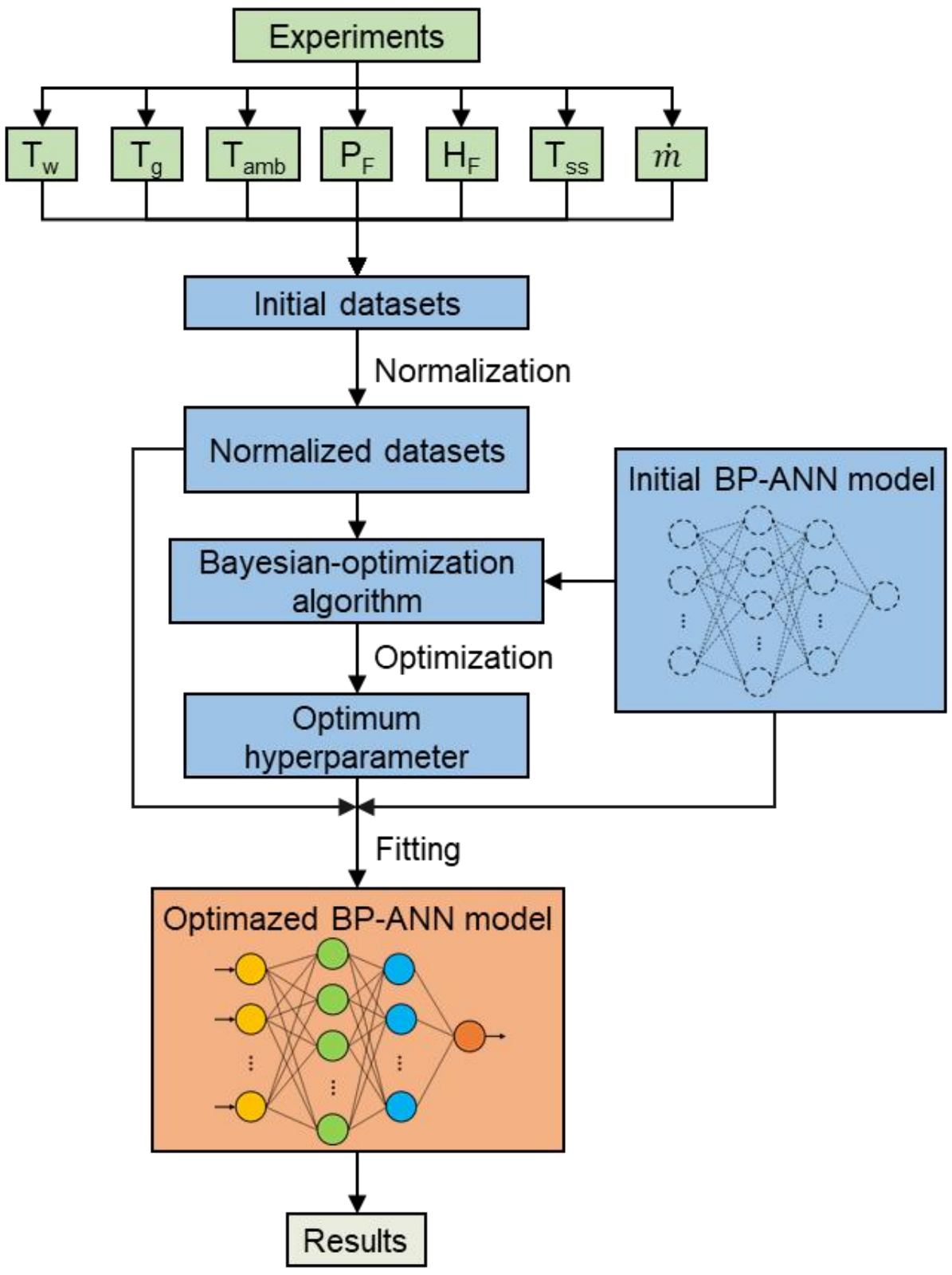

Fig. 4 The flow chart of using BP-ANN for productivity prediction of the STD system.

The BO process for BP-ANN is illustrated in Fig. 5. The optimization process for RF with BO is similar. The initial BP-ANN model and normalized datasets are inputted to construct the adjustment function as

$$R^2 = BPNN(h_1,\ h_2,\ h_3,\ h_4,\ h_5,\ D) \tag{10}$$

where $R^2$ is the coefficient of determination, $h_1 \sim h_5$ are the hyperparameters of BP-ANN, and D is the normalized dataset, which is constant in the fitting process.

The purpose of BO in this work is to find the optimum hyperparameters that can obtain the maximum $R^2$. BO is especially suitable for "black box" function optimization problems, that is, the dimension is relatively high, and the function value is difficult to obtain[50]. Firstly, the computer generates the random data points of the adjustment function within the defined domain. Based on the probabilistic surrogate model, the prediction function and confidence interval for the adjustment function are established. The acquisition function is employed to predict the quasi-optimal hyperparameters. If the threshold, defined as the maximum number of iterations in this work, is not met, the $R^2$ is calculated using the quasi-optimal hyperparameters. A new data point is generated based on this $R^2$ value and added to the existing data points of the adjustment function. The process is then repeated. Once the threshold is met, the quasi-optimal hyperparameters are considered the final optimum hyperparameters and are outputted. In this work, the probabilistic surrogate model is based on Gaussian process regression, while the acquisition function is based on the upper confidence bound. Details in "Supporting Information S7".

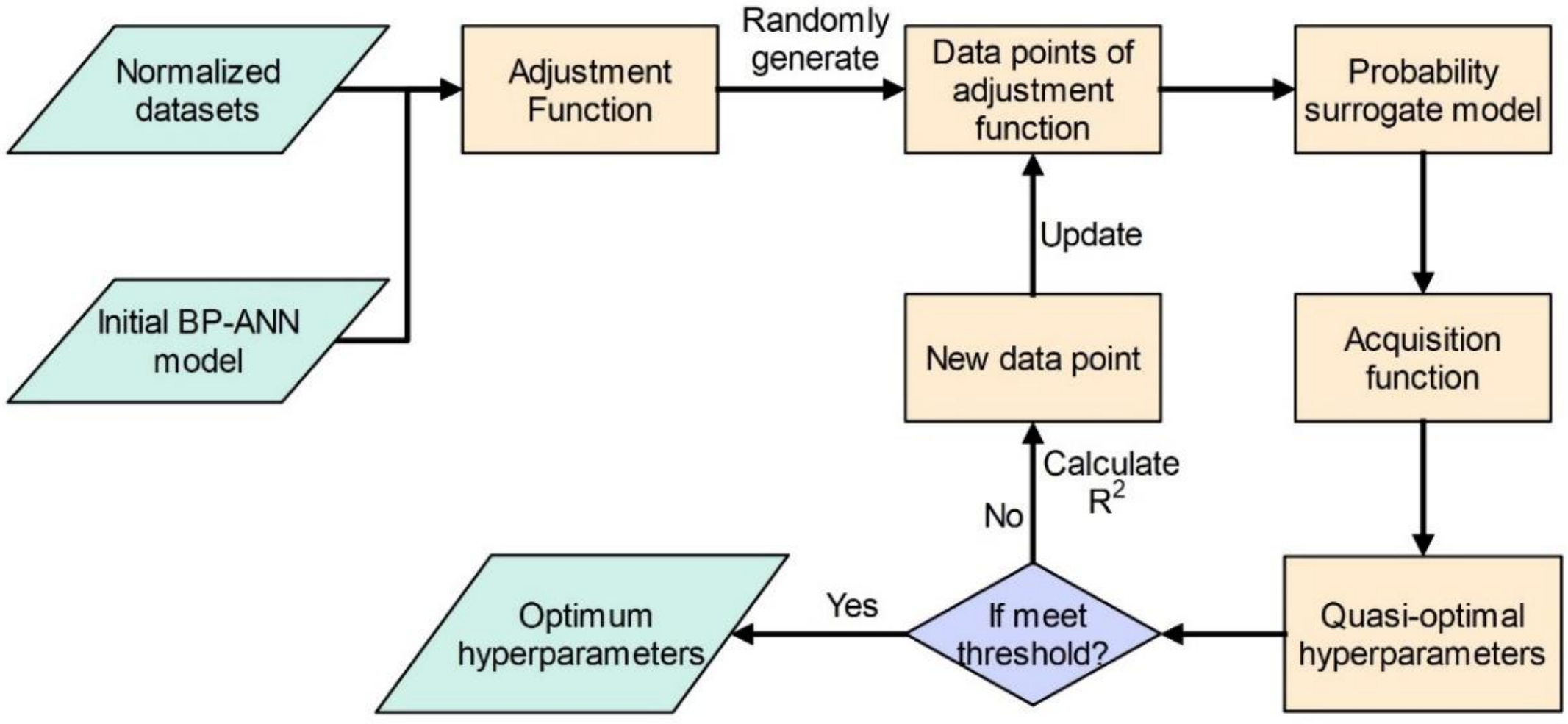

Fig. 5 Schematic diagram of the Bayesian optimization process.

### (3) RF model

In recent years, the RF algorithm has been one of the most popular algorithms and has been widely used in Kaggle Competitions and academic dataset analysis [51], as well as in the field of STD [19]. RF is an ensemble learning method, which integrates multiple decision tree (DT) models. Based on the bagging, datasets are divided into numerous bootstrap samples to fit the DT models. The predicted result of RF is obtained by averaging the results of the DT models. The DT models of RF are independent of each other and can be calculated in parallel. Therefore, RF usually has a higher model training speed when dealing with large-scale datasets [52].

There are two main functions of RF for analyzing the STD system, including predicting the productivity and ranking the importance of influence factors:

a) Productivity prediction

RF is constructed by classification and regression trees (CART), which provides a flexible and powerful algorithm for regression tasks, offering interpretability, robustness, and handling of nonlinear relationships and mixed data types [53]. Utilizing the CART algorithm, a binary DT is constructed to partition each dimension into two regions. The output values are obtained within each region of the tree. Based on the heuristic algorithm, one sample, such as the $j_{\text{th}}$ sample, will be chosen as the slicing variable and slicing point, which defines two regions, $R_1(j)$ and $R_2(j)$

$$R_1(j)=\{X_{jq}|y\leqslant y_j\} \quad R_2(j)=\{X_{jq}|y>y_j\} \tag{11}$$

where, $q$ is the dimension label, which is an integer in the range $[1,\mathrm{k}]$, and $\mathrm{k}$ is the number of the independent variables. $\mathrm{X},\mathrm{y}$ are the input variables. In this work, the input dataset X is an array containing experimental independent variables such as $T_w$, $T_g$, $T_{amb}$, $P_F$, $H_F$, and $T_{ss}$. $\mathrm{y}$ is the corresponding productivity, $\dot{m}$. The DT uses the principle of minimizing the squared error

$$MIN = \min_j\left[\min_{c_1}\sum_{x_i\in R_1(j)}(y_i-c_1)^2+\min_{c_2}\sum_{x_i\in R_2(j)}(y_i-c_2)^2\right] \tag{12}$$

where $c_1$ and $c_2$ are the average values of y in $R_1(j)$ and $R_2(j)$, respectively.

Traversing the variable $j$, the optimal segmentation point can be obtained for fixed input variables. Repeating the above process $N-1$ times, the input space can be divided into $N^k$ regions. The average output value, $o_M$, of each region is

$$o_M=\frac{1}{N_M}\sum_{y_j\in R_M(j)}y_i \quad, M=1,2,\cdots,N^k \tag{13}$$

where $M$ is the label of the regions; $R_M(j)$ is the region of label $M$; $N_M$ is the number of elements in the region $M$. The output of the decision tree model is

$$f(X_i)=o_M \quad, \quad X_i\in R_M(j) \tag{14}$$

The output $\mathrm{y}^{RF}$ of the RF model is

$$y^{RF}=F(X_i)=\frac{1}{n}\sum_{k=1}^{n}f_k(X_i) \tag{15}$$

where n is the number of DTs in the RF model.

After fitting the RF model, the prediction model $\mathrm{F}(\mathrm{X}_i)$ is obtained. If an input dataset X' is provided, the predicted productivity y' can be expressed as

$$y' = F(X') \tag{16}$$

b) Importance ranking

The quantitative importance of feature factors is calculated by using the Gini impurity. Based on the binary decision tree, the Gini impurity is

$$GI_e = 2\hat{p}_e\left(1 - \hat{p}_e\right) \tag{17}$$

where $\hat{p}_e$ is the estimated probability that the sample belongs to any class at node $e$.

The importance of variable $X_i$ at node $e$, $\mathrm{VIM}_{i\,e}^{DT}$, that is, the change in Gini impurity before and after the branch of node $e$ is

$$VIM_{i\,e}^{DT} = GI_e - GI_l - GI_r \tag{18}$$

where $GI_l$ and $GI_r$ is the Gini impurity of the two new nodes split by node $e$.

By calculating all the $\mathrm{VIM}_i^{DT}$ about variable $X_i$, the importance of the variable $X_i$ in the random forest can be obtained. The specific calculation processes can be found in "Supplementary Information S3". The overall flow chart of using RF for predicting productivity and calculating the factor importance (weighted value) of the STD system is similar to that of BP-ANN, as shown in "Supporting Information Fig. S1 and Fig. S2".

To evaluate the prediction accuracy, three indicators were used in this work, namely relative prediction error ($\delta$), mean relative prediction error ($\overline{\delta}$), and the coefficient of determination ($R^2$). The definitions are shown in Table 3, which are the most typical evaluation metrics in STD field. In table 3, $y_i$ is the experimental productivity of dataset i, $f_i$ is the predicted productivity of dataset i, $\bar{y}$ is the average productivity of all calculated datasets. Meanwhile, the calculation used the 5-fold cross-validation as discussed in "Supporting Information S8". The fitting process with the optimal hyperparameter was performed 10 times with different training and testing datasets that were split randomly. The output evaluation indicators are the mean of these 10 results. Although this method may slightly reduce the accuracy of the fitted machine learning model, it mitigates the overfitting problem and significantly enhances the model's generalization ability.

Table 3 The definitions of main evaluation indicators

| Evaluation indicators | Expression |
| --- | --- |
| Relative prediction error ($\delta$) | $\delta = \frac{y_i - f_i}{y_i} \times 100\%$ |
| Mean relative prediction error ($\overline{\delta}$) | $\overline{\delta} = \frac{1}{n}\sum_{i=1}^{n}\delta$ |
| Coefficient of determination ($R^2$) | $R^2 = 1 - \frac{\sum_{i=1}^{n}(y_i - f_i)^2}{\sum_{i=1}^{n}(y_i - \overline{y})^2}$ |

## 3. Results and discussions

Besides optimizing the data collection process for better analysis. The dataset characteristics also should be considered, which may have significant effects on the interdiscipline between ML and STD. However, it was overlooked in previous works. Therefore, dataset characteristics should be considered in the whole process of interdisciplinary study, making the process more general and standard, which is crucial for promoting ML to be a general tool for analyzing STD. Fig.6 illustrates the optimized process flow of the interdisciplinary study proposed by this work. The optimized process flow consists of seven steps:

(1) First, the main purpose of the study should be determined. Currently, the main applications of ML include predicting values, searching extreme values, and analyzing influence factors. Other applications might also be explored in the future.

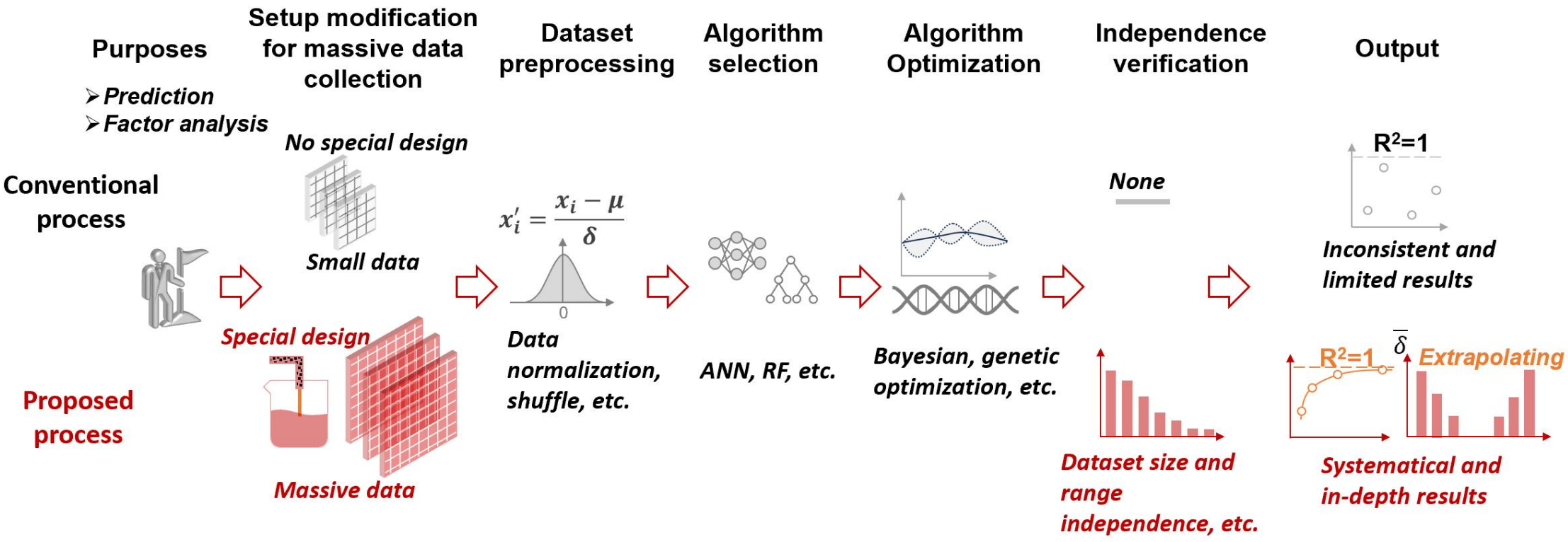

Fig.6 Process flow for the interdisciplinary study between machine learning and solar-thermal desalination. The proposed new process flow will enable more reliable, systematic, and in-depth analyses compared to the conventional process flow in previous studies.

(2) Secondly, it is suggested to establish an experimental platform that is designed especially for massive dataset collection, which is the key step. Generally, in the previous studies, the platforms weren't intentionally optimized, thus usually a quite small dataset (< 200) was collected due to the limitation of the conventional setup.

(3) The datasets, obtained from the designed STD platform, are preprocessed, such as normalization, shuffle, removal of invalid datasets, and so on.

(4) Algorithm selection, which is based on the purpose of the study and the characteristics of datasets.

(5) The algorithm should be optimized by using optimizers for choosing proper hyperparameters, such as "max_depth", "max_features", and "n_estimators" in RF. The optimizers include Bayesian, genetic algorithm, Harris Hawks Optimizer, etc.

(6) Importantly, "independence verification" of dataset characteristics should be carried out, which is missing in previous studies on STD systems. To select the best algorithm and give a consistent result, the dependence of dataset characteristics, such as size, range, factors, etc. should be checked carefully. In the review of previous works, there are some inconsistent conclusions, partially due to missing the step of independence verification. It is easier to obtain valuable and universal conclusions with independence verification.

(7) Finally, the desired results are outputted. The conventional process only provides a single value of $R^2$ or other simple indicators. More meaningful results can’t be obtained, such as productivity extrapolation.

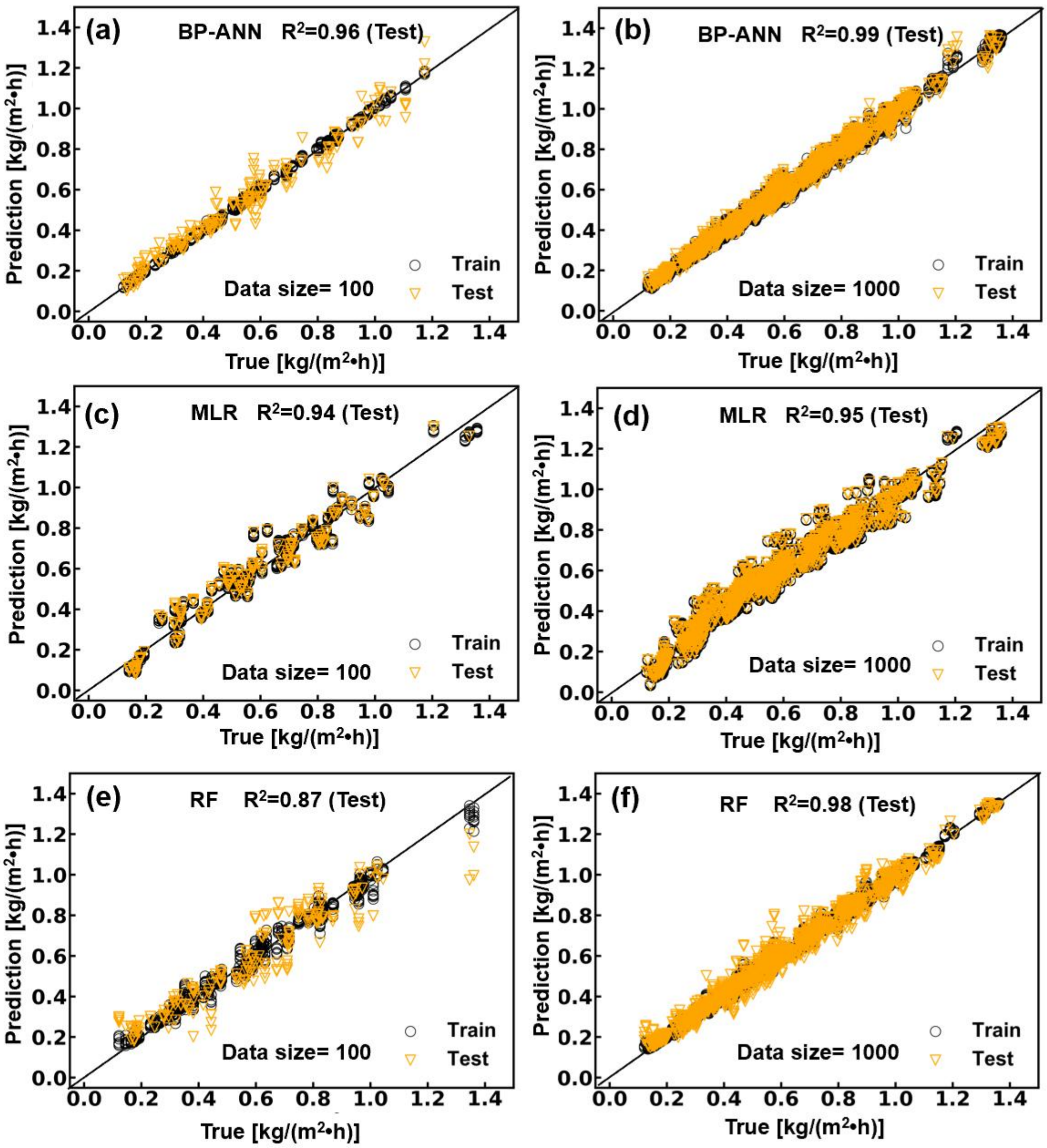

Fig. 7 Prediction and the true value of productivity by using (a) BP-ANN and 100 data points, (b)BP-ANN and 1,000 data points, (c) MLR and 100 data points, and (d) MLR and 1,000 data points, (e)RF and 100 data, (f) RF and 1,000 data. The $R^2$ in the figure is the value of the testing set.

Based on the optimized data collection process and interdisciplinary process flow, the effect of dataset size on the prediction of productivity by using BP-ANN, MLR, and RF was first investigated. A series of datasets were randomly selected from the entire dataset to represent different dataset sizes. The results are shown in Fig.7. The $R^2$ of the testing set in BP-ANN is as high as 0.96 based on 100 datasets (Fig. 7a). The $R^2$ of the testing set increases from 0.96 to 0.99 when the dataset size increases from 100 to 1,000, which indicates great predicting accuracy of BP-ANN under large dataset size (Fig. 7b). On the other hand, the $R^2$ of the testing set of MLR slightly increases from 0.94 to 0.95 when the dataset size increases from 100 to 1,000 (Fig. 7c and 7d). The $R^2$ of RF shows the most significant improvement with the increases in the dataset size. The $R^2$ of the testing process is as low as 0.87 for 100 datasets and as high as 0.98 for 1,000 datasets, which is comparable to BP-ANN. (Fig. 7e and 7f).

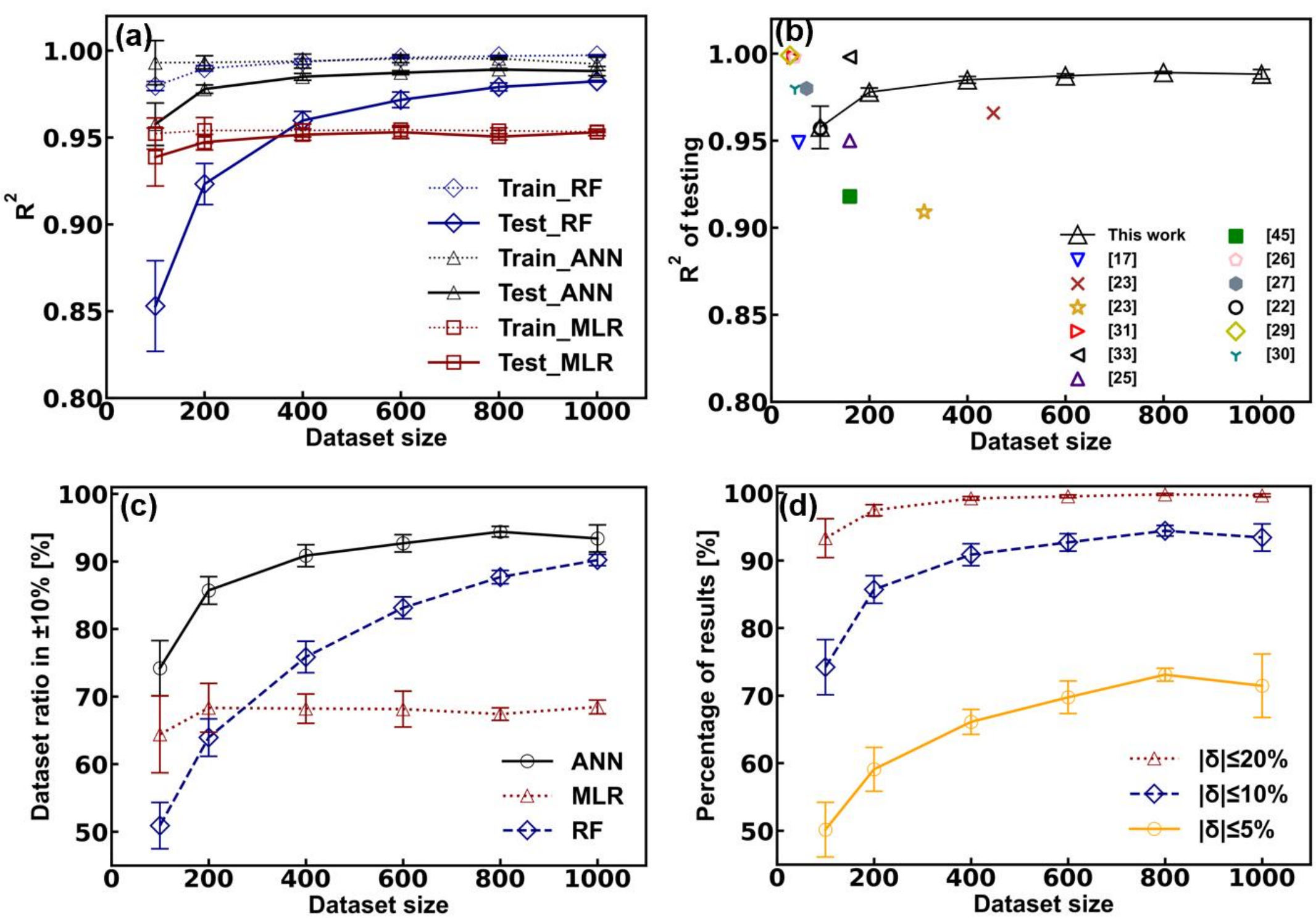

Fig. 8 The results of predicting the productivity of solar stills using different algorithms. (a) $R^2$ value for the training sets and testing sets of BP-ANN, RF, and MLR based on varying dataset sizes. (b) $R^2$ value for the testing set of BP-ANN in various works. (c) Percentage of productivity in the testing sets for which $|\delta| \leq 10\%$, using BP-ANN, MLR, and RF. (d) Percentage of productivity in the testing sets for different ranges of $\delta$, using BP-ANN.

As shown in Fig.8, the results show that their performance varies a lot when the dataset size ranges from one hundred to one thousand. The $R^2$ of three algorithms under different dataset sizes are summarized in Fig. 8a. For BP-ANN and MLR, the $R^2$ of the testing sets (more important than training sets) slightly increases with the dataset size. However, it dramatically increases for RF. In general, BP-ANN has a better performance than the others. However, the performance of MLR and RF, superiority strongly depends on the dataset size. RF will be better than MLR when the dataset size is larger than 400.

Therefore, the choice of the algorithm should consider the effect of dataset size on the accuracy. The results with a small dataset make an incomplete conclusion. Some previous studies with a dataset size smaller than one hundred cannot take a comprehensive consideration.

Although BP-ANN outperforms RF in accuracy, the mean fitting time of BP-ANN is 5.9 times longer than that of RF. This is because RF consists of many independent decision trees, which can be computed in parallel, hence the increased computational speed. Therefore, given the training speed, RF might be a better choice when the dataset size is much larger than 1,000. Thus, the interdisciplinary research of ML and STD should consider the influence of dataset size to reach more generalizable conclusions.

In comparison to the clear trends of dataset size effect found in this work, the results collected from the references are quite inconsistent as shown in Fig. 8b. This is because different works have different optimization models, working conditions, analyzing standards, and so on, which makes it difficult to compare with each other. Therefore, a successful and comprehensive investigation of the STD system by ML relies on a consistent analysis and sufficient dataset, which is difficult to obtain or conclude by collecting the results of current references. This emphasizes the importance of designing a rational experimental system for systematical dataset collection, as well as a standard process flow for ML analysis.

Fig. 8c shows the percentage of predicted productivity that are within 10% of error, i.e., $|\delta| \leqslant 10\%$, by using BP-ANN, MLR, and RF. For BP-ANN, the percentages of δ within ±10% are 74.2%, 90.8%, and 93.4% for 100, 400, and 1000 datasets, respectively. Besides, 70% and 99.6% of the δ are less than ±5% and ±20%, respectively, when the dataset size is more than 600 (Fig. 8d). In contrast, by using MLR, only 60% to 70% of the δ are less than ±10%, even with a dataset size as high as 1,000. On the other hand, for RF, as the dataset size increases from 100 to 1,000, the percentage of results for δ within ±10% rises from 50.9% to 90.2%. This is comparable to BP-ANN when the dataset size reaches 1,000, which is similar to the performance based on $R^2$.

The results show that the performance of prediction might vary a lot although the dataset size only ranges from several hundred to one thousand. Therefore, the choice of the algorithm should consider the dataset size and its effect on the accuracy, time cost, and other factors of each algorithm. Without a comprehensive consideration, previous studies with a small dataset size may suggest that BP-ANN outperforms RF, but the impact analysis of dataset size indicates that RF may be more time-effective when the dataset size exceeds one thousand. Thus, the interdisciplinary research of ML and STD should fully consider the influence of dataset size to reach more generalizable conclusions.

Secondly, besides the dataset size, the dataset range is another factor that affects the interdiscipline between ML and STD. Great dataset range indicates that the dataset covers all the possible conditions in practical applications, such as a wide water temperature or ambient temperature range. Herein, the factor importance (weighted value), which could guide the system optimization by selecting the important factors of the desalination system for future optimization [18], is taken as an example of studying the effect of dataset range.

The factor importance is obtained by analyzing the connection between productivity and factors by using the RF algorithm. Fig. 9a shows the importance of $T_w$, $T_g$, $T_{amb}$, $P_F$, $T_{ss}$, and $H_F$ in predicting productivity. The importance of the $T_w$, $T_g$, $T_{amb}$, $P_F$, and $T_{ss}$ is relatively stable in different dataset sizes. This might be because the dataset range is the same for different dataset sizes. As aforementioned, different dataset sizes are obtained by randomly selecting a given amount of dataset from the entire dataset. Thus, every dataset covers almost all the experimental condition ranges, such as temperatures, fan powers, and so on. The difference is mainly the data density in each condition.

On the contrary, the importance of the random list ($R_L$) decreases fast as the dataset size increases. $R_L$ is a random list composed of 1, 2, and 3. The importance of $R_L$ should be 0 in an ideal calculation. However, a subtle connection between $R_L$ and productivity would exist when the dataset size is finite. The importance of $H_F$ is similar to that of $R_L$, which indicates that $H_F$ doesn't affect productivity. For small dataset sizes, such as 100, it is difficult to rank $T_{ss}$, $H_F$, and $R_F$ reliably. Therefore, analyzing the factor importance with different dataset sizes could help us to clearly distinguish the correlation factors, which is of great importance in complex systems with many influence factors.

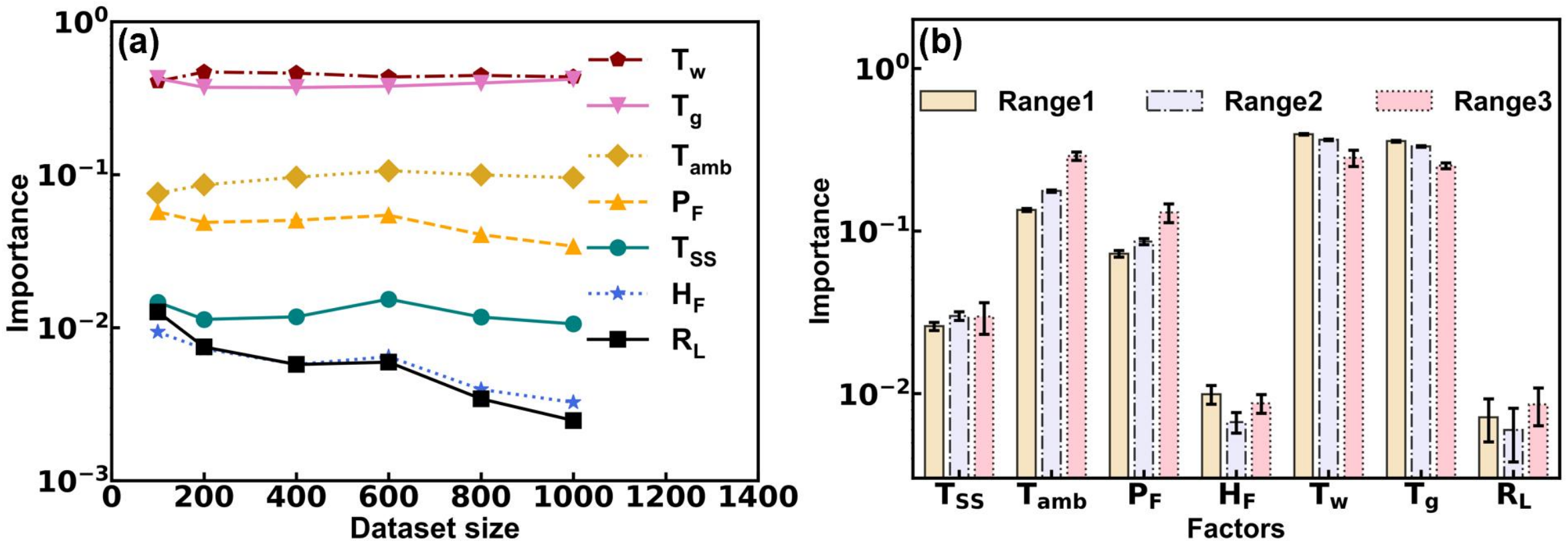

Fig. 9 The quantified factor importance (weighted value). (a) For different dataset sizes, (b) For different ranges of $T_w$, where Range 1 is $T_w$ from 30 to 85°C), Range 2 is $T_w$ from 40 to 75°C), and Range 3 is $T_w$ from 50 to 65°C. $R_L$ is a random list for comparison.

The effect of the water temperature range is investigated as an example of showing the effect of dataset range. Herein, three water temperature ranges are selected, including Range 1 ($T_w$ = 30 to 85°C), Range 2 ($T_w$ = 40 to 75°C), and Range 3 ($T_w$ = 50 to 65°C). Fig. 9b shows the factor importance of different ranges. In the case of Range 1, the rank of importance is $T_w > T_g > T_{amb} > P_F > T_{ss} > H_F > R_L$. The importance of $T_{amb}$ and $P_F$ increases gradually and the importance of $T_w$ and $T_g$ decreases gradually as the range of $T_w$ becomes more and more narrow. As compared to Range 1, in the case of Range 3, the importance of $T_{amb}$ increased by 115%. It becomes the most important factor, slightly higher than $T_w$ and $T_g$. This means that if the possible range of an important factor is not completely measured, the importance of this factor will be significantly underestimated, while the importance of other factors will be overestimated. The effect of the data ergodicity from the aspect of $P_F$ is shown in Supporting Information Fig. S4, which shows that the convergence range doesn't cover up or mislead the importance of the factors. Therefore, a great dataset range is quite important.

In addition, the effect of dataset range on the model generalization performance is investigated. The model generalization performance means the ability of the model to extrapolate productivity, i.e., to predict the productivity of conditions that are beyond the existing experimental dataset range. Herein, the model generalization performance of BP-ANN is investigated due to its high accuracy among the three algorithms. The dataset is divided into 11 parts according to the range of $T_w$, as shown in Fig. 10a. Based on the range of $T_w$, four cases are investigated, namely "Case 1" to "Case 4". In each case, 500 datasets are selected randomly for training and testing. For example, in "Case 1", 500 datasets from range 2 to range 10 (yellow shadow) are selected for training and testing while the datasets in range 1 and range 11 (green rectangle) are used for prediction, which can validate the extrapolation model. In "Case 4", 500 datasets from range 5 to range 7 are selected for training and testing, and the rest datasets are used for validating the extrapolation model. After training and testing, an ANN model is established, which can be used to calculate productivity in the extrapolation range. In this section, all the predicted datasets are out of the training and testing range, which demonstrates the model generalization performance.

To demonstrate the accuracy of extrapolation, the mean relative prediction error ($\overline{\delta}$) was calculated for all extrapolation ranges, as shown in Fig. 10b and Table 4. A low $\overline{\delta}$ indicates that most extrapolated productivities are close to the experimental productivities, hence a high extrapolation accuracy. The results show that a similar trend can be observed in all cases. $\overline{\delta}$ becomes larger when the predicted ranges are farther away from the edge of the training and testing dataset. Meanwhile, $\overline{\delta}$ increases more significantly in low water temperature compared to high water temperature. $\overline{\delta}$ is only around 4% to 5% when the predicted range is adjacent (0 - 5°C of difference) to the upper edge of training and testing sets. $\overline{\delta}$ remains as low as 9.1% even when the $T_w$ in the predicting set is 20 to 25°C higher than the $T_w$ in the training and testing set. On the contrary, $\overline{\delta}$ will be 8.6% to 13.2% for datasets adjacent to the lower edge of training and testing and larger than 15% when $T_w$ is further away from the lower $T_w$ edge of training and testing. In general, it can be inferred that BP-ANN might be used to predict the productivity of the unmeasured conditions that are adjacent to the measured conditions, but the accuracy might be different near the upper edge and lower edge.

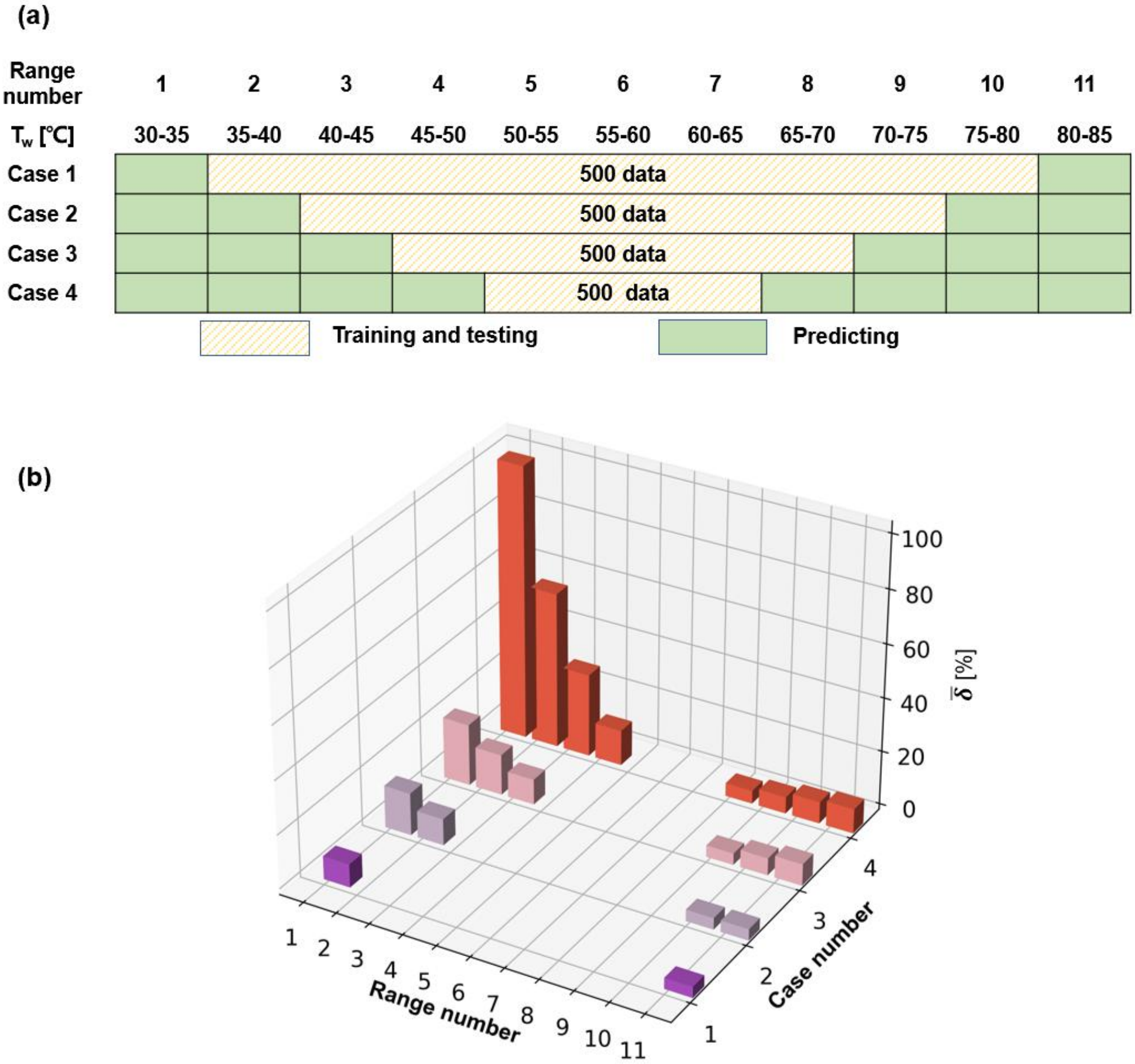

Fig.10 (a) Definition of cases and ranges. 500 datasets are randomly selected for training and testing in each case. (b)The average relative error ($\overline{\delta}$) of productivity extrapolation under different ranges of $T_w$ (Case 1 to Case 4).

Table 4 The mean relative prediction error ($\overline{\delta}$) of productivity extrapolation.

| Range number | 1 (30-35℃) | 2 (35-40℃) | 3 (40-45℃) | 4 (45-50℃) | 8 (65-70℃) | 9 (70-75℃) | 10 (75-80℃) | 11 (80-85℃) |
|---|---|---|---|---|---|---|---|---|
| Case 1 | 8.6±0.8% | - | - | - | - | - | - | 4.0±0.5% |
| Case 2 | 16±2% | 9.8±0.7% | - | - | - | - | 4.2±0.5% | 4±1% |
| Case 3 | 23±3% | 15±2% | 9±1% | - | - | 4.3±0.2% | 6.3±0.7% | 8±1% |
| Case 4 | 100±40% | 60±30% | 30±10% | 13±4% | 4.9±0.3% | 6.2±0.7% | 8±2% | 9±5% |

## 4. Prospects

This work presents an example of interdisciplinary research between ML and STD, transcending the limitations of conventional STD studies that rely solely on productivity data fitting. The methodology and process have broad applicability and can be extended to various STD systems with different configurations or components, such as contactless desalination design [54], multistage design[37], solar still with condensers [55], or even humidification-dehumidification systems [56], owing to their shared operational processes: water heating, evaporation, and condensation. Furthermore, this approach could potentially be applied to other scientific fields as well.

Fig. 11 summarizes the flowchart for different systems. Conventional studies often exhibit diverse flowcharts across different works. In this work, apart from the conventional process, it is suggested to optimize the system for collecting more data first, instead of directly collecting data from conventional systems. For example, optimizing the water collection part. In this case, the minimum time step for data collection should be investigated first. Later, the key limitations of obtaining shorter time steps should be revealed, which might be a challenge. The limitations may vary with the system and should be investigated thoroughly. Future work should pay more attention to optimizing the system for more data.

Besides, mining more scientific findings through interdisciplinary studies between ML and STD, as well as extending their real-world applications, would be very valuable directions. For example, integrating machine learning and data acquisition processes to construct prediction models much faster and more accurately [57], and obtaining the best real-time operational parameter combinations for different STD systems. The proposed process in this work provides a new approach to realizing more valuable applications.

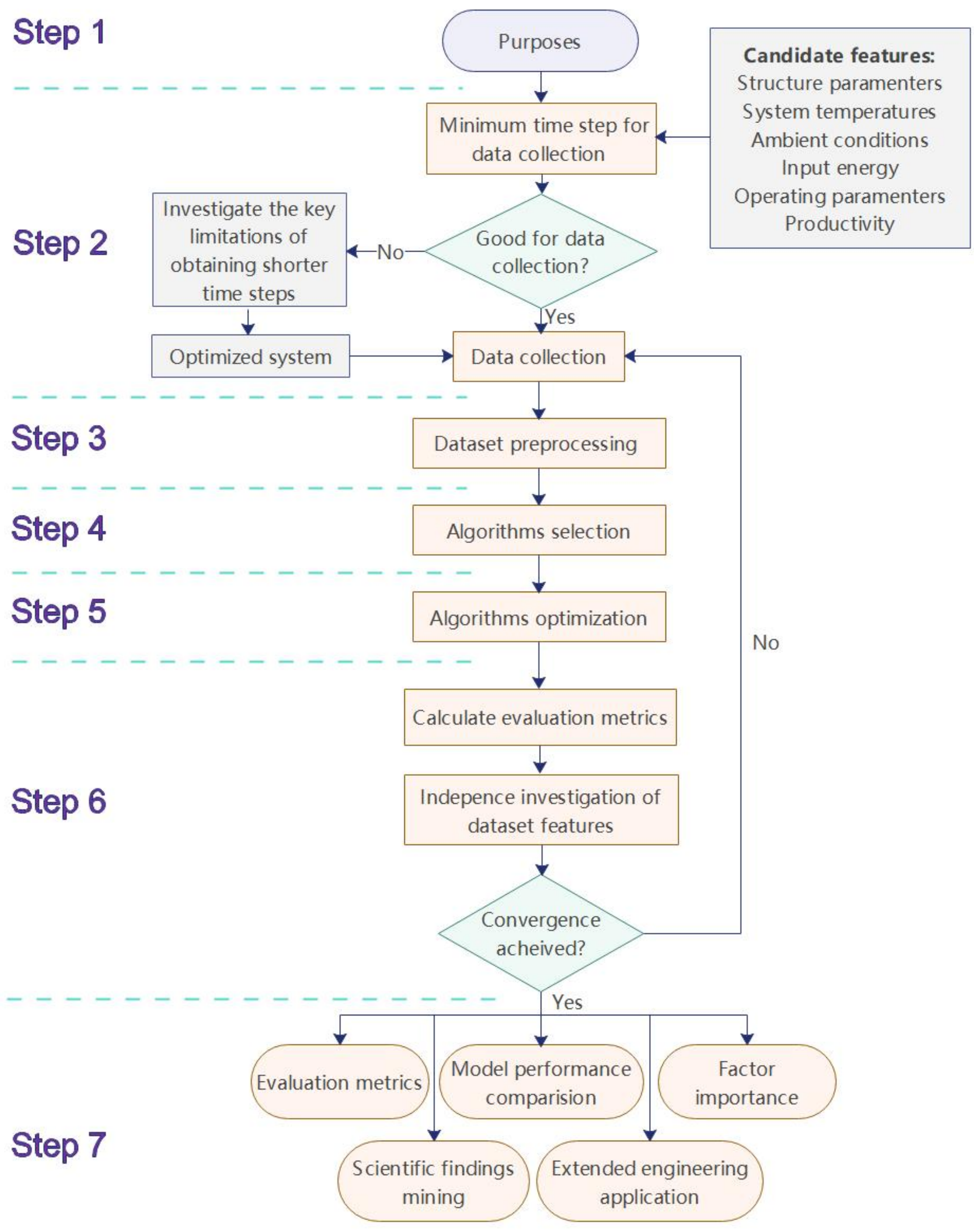

Fig. 11 Flowchart of the interdisciplinary for solar-thermal desalination systems.

## 5. Conclusion

In conclusion, this study tackles the challenges inherent in the interdisciplinary integration of solar thermal desalination and machine learning, particularly the limited data volumes, insufficient analytical depth, and inconsistent results. To overcome these limitations, a standard process is proposed, including seven steps, which are different from the conventional process in three aspects.

Firstly, it is emphasized that optimizing data acquisition processes is essential for enhancing data availability before applying machine learning techniques. For example, by refining the water collection process in solar stills, the data collection time was reduced by 83.3%, resulting in a dataset of over 1,000 samples—far exceeding the volumes reported in previous studies.

Secondly, independence validation for dataset characteristics is recommended when assessing and comparing the performance of different machine learning models. The investigation of Multiple Linear Regression, Random Forest, and Artificial Neural Network models highlights the strong correlation between dataset size and the predictive accuracy of these models for solar still productivity. To ensure consistency and systematicity in research outcomes, it is imperative to evaluate model performance across varying dataset sizes, data ranges, and other relevant characteristics.

Lastly, the results in this work highlight the potential of integrating solar thermal desalination with machine learning beyond simple data fitting. By utilizing larger datasets and adhering to a rigorous research process, it becomes more feasible to uncover novel scientific insights and expand engineering applications. This includes extrapolating productivity, analyzing factor importance, constructing prediction models much faster and more accurately, or implementing real-time optimization of operational parameters in the future, thereby advancing the field of solar thermal desalination and promoting interdisciplinary research.

## 6. Conflicts of interest

There are no conflicts of interest to declare.

## 7. Acknowledgment

The work was sponsored by the National Key Research and Development Program of China (2018YFE0127800), Science, Technology & Innovation Funding Authority (STIFA), Egypt and China, under the grant (40517). The work was carried out at the National Supercomputing Center in Tianjin (NSCC-TJ), and the calculations were performed on TianHe-HPC. We also thank Hua Bao for a valuable discussion.